# Kink solitons in DNA

S. Zdravković . M. V. Satarić . M. Daniel

S. Zdravković
Institut za nuklearne nauke Vinča,
Univerzitet u Beogradu, Poštanski fah 522,
11001 Beograd, Serbia
szdjidji@vinca.rs

M. V. Satarić
Fakultet tehničkih nauka, Univerzitet u Novom Sadu,
21000 Novi Sad, Serbia
bomisat@neobee.net

M. Daniel
Centre for Nonlinear Dynamics,
School of Physics, Bharathidasan University,
Tiruchirappalli  620 024, India
daniel@cnld.bdu.ac.in

**Abstract**   We here examine the nonlinear dynamics of artificial homogeneous DNA chain relying on the plain-base rotator model. It is shown that such dynamics can exhibit kink and antikink solitons of sine-Gordon type.
   In that respect we propose possible experimental assays based on single molecule micromanipulation techniques. The aim of these experiments is to excite the rotational waves and to determine their speeds along excited DNA. We propose that  these experiments should be conducted either for the case of double stranded (DS) or single stranded (SS) DNA. A key question is to compare the corresponding velocities of the rotational waves indicating which  one is bigger. The ratio of these velocities appears to be related with the sign of the model parameter representing ratio of the hydrogen-bonding and the covalent-bonding interaction within the considered DNA chain.

Introduction

It is well known that DS-DNA molecule represents a double helix with a helical pitch of about 11 nucleotide pairs per turn. Each nucleotide consists of sugar, phosphate and base. All the sugars and the phosphates are the same which means that a sequence of base pairs represents a piece of information.
   We here restrict to the case of homogeneous DS-DNA chain avoiding the impact of inhomogeneities and we study the possible excitation and the propagation of rotational kinks.
   Special importance for biological systems play the weak interactions which are comparable with physiological thermal energy. A representative example is the hydrogen bond, connecting different strands in DS-DNA molecule. These weak interactions are mostly responsible for nonlinear effects in biomolecules, including DNA, of course. This nonlinearity causes that crucial equations, corresponding to different models describing DNA dynamics, are nonlinear partial differential equations. Among the solutions of such equations especially important are waves called solitons. The essential feature of these waves is their robustness and stability, which is biologically extremely important.

Nonlinear physics of DNA had begun some thirty years ago [1]. A couple of models describing some aspects of nonlinear dynamics of DNA have been developed so far [2]. The corresponding basic equations are Boussinesq, sine-Gordon and nonlinear Schrödinger equation. Their fundamental solutions are bell shaped solitons [3,4], kink (antikink) stable waves [5-8], and breather solitons [9-11], respectively. In what follows we focus on the sine-Gordon model in the context of possible experimental proposal aimed to verify its validity at least under specific in vitro conditions.

According to the plane-base rotator (PBR) model, the main attention has been paid to base rotations in the plane perpendicular to the helical axis around the backbone structure [7,8]. The first model was proposed by Yomosa [12,13] and improved by Homma and Takeno [14]. Further improvement was done recently by Daniel and Vasumathi [7,8] by introducing an analogy between the PBR model and the Heisenberg's spin model for ferromagnetic chain, and this is what we strongly rely on in this paper.

The main objective of this paper is to suggest how the solitonic speed and its width could be determined experimentally, in the context of single molecule manipulation. If we properly estimated these two values we would know the ratio of two key parameters describing the transverse and the longitudinal interaction of nucleotides in DS-DNA. We will see that two distinct wave velocities are of clue relevance. These are the kink velocity in DS-DNA and the plane rotational wave speed in SS-DNA. By comparing these velocities one is able to determine the sign and the value of the parameter describing the weak hydrogen bond interaction, as well as the form of equation governing the pertaining dynamics of DNA chain.

The suggested experiment proposals are based on micromanipulation techniques. During the past two decades it is getting possible to mechanically manipulate a single molecule. This has become a very attractive research field for a number of scientists and the very first paper [15] has represented a base for a lot of works studying various elastic properties of different molecules [16-28]. Those experiments were accompanied by theoretical modeling [29-32], numerical simulations [33,34], experimental proposals [35-37] and review papers [38-40].

The paper is organized as follows. We start with very brief descriptions of the PBR model and the Heisenberg spin model, respectively. More detailed elaborations of this can be found in papers [7,14,41] and in references therein. It may be of interest to point out that this approach can be also used to study modulation instability in DS-DNA [42] as well as DNA-protein interactions [43].

Then, we offer the experimental proposal to determine the speeds of rotating waves in both DS-DNA and SS-DNA homogeneous chains. Also, we give a comment on what can be thought of by the width of the kink-soliton in relation with the experiment. Finally, we close the paper with some concluding remarks.

**PBR model for DNA dynamics**

Let z be a coordinate axes along the helix of DS-DNA chain. According to the PBR model we can imagine all the sugars and the phosphates of the nucleotides posed on the surface of a cylinder while the bases are alike thin rods oriented towards inside [7]. Figure 1 shows two nucleotides at a same position belonging to different strands. Hence, small circuits represent phosphate-sugar part of the nucleotides while arrows stand for radius vectors of tips of the bases.

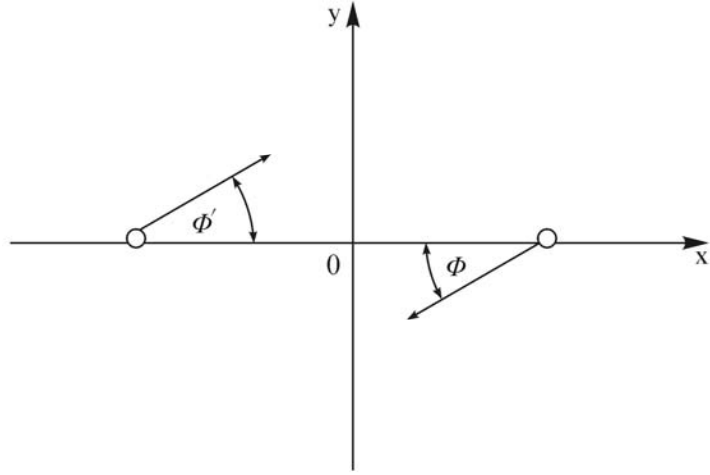

**Fig. 1** Oscillation of two complementary bases in xy-plane

There are four degrees of freedom for base vectors. These are polar angles $\theta_n$ and $\theta'_n$ and corresponding azimuthally angles $\Phi_n$ and $\Phi'_n$. This means that movement of the sugar-phosphate parts of the nucleotides is ignored, while the rotational oscillations of the bases dominate the dynamics. The index prime has been introduced to distinguish the two strands.

In order to probe the simplest version of the PBR model we suggest the following topology of homogeneous DS-DNA chain. By appropriate single molecule manipulation the chain should be completely unwound thus removing its helicity (Fig. 2). Under such circumstances we can safely take $\theta_n = \theta'_n = \pi/2$ since the base vectors are all in xy-plane. Hence, Fig. 1 shows that we are concerned just about the azimuthal angles $\Phi_n$ and $\Phi'_n$.

Let $(S_n^x, S_n^y, S_n^z)$ and $(S_n'^x, S_n'^y, S_n'^z)$ be the components of the radius vectors shown in Fig. 1. It is obvious that $S_n^x = -\cos\Phi_n$, $S_n'^x = \cos\Phi'_n$, etc. for $\theta_n = \theta'_n = \pi/2$. If the angles $\theta_n$ and $\theta'_n$ are arbitrary then

$$S_n^x = -\sin\theta_n \cos\Phi_n, \quad S_n^y = -\sin\theta_n \sin\Phi_n, \quad S_n^z = \cos\theta_n \tag{1}$$

$$S'^x_n = \sin\theta'_n \cos\Phi'_n, \qquad S'^y_n = \sin\theta'_n \sin\Phi'_n, \qquad S'^z_n = \cos\theta'_n. \tag{2}$$

Of course, the unit lengths of these vectors are assumed. Notice that

$$S^x_n S'^x_n < 0, \qquad S^y_n S'^y_n < 0, \qquad S^z_n S'^z_n > 0. \tag{3}$$

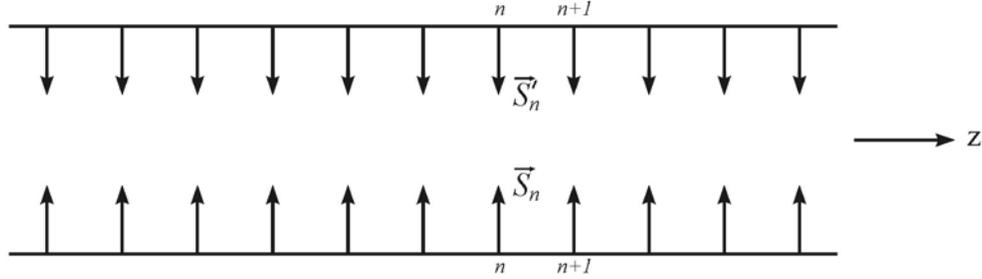

**Fig. 2** Two chain spin model or spin ladder

The expression for a distance between the tips of the bases of the same pair was derived [7]. A term existing in the formula for the square of this distance, relevant for this paper, is proportional to the following trinome

$$T = S^x_n S'^x_n + S^y_n S'^y_n - S^z_n S'^z_n. \tag{4}$$

According to Eq. (3), we can see that all the terms in Eq. 4 have the same sign.

**Heisenberg spin model**

Spin excitations in ferromagnets can be expressed in terms of the Heisenberg's nearest neighbour spin-spin exchange interaction. For the simplest isotropic case, the Hamiltonian for the one-dimensional ferromagnet is

$$H = -J\sum_{ij} \vec{S}_i \cdot \vec{S}_j \tag{5}$$

where $\vec{S}_i = (S^x_i, S^y_i, S^z_i)$ is the spin operator and $J$ is the exchange integral [44]. This Hamiltonian can be generalized. For example, if there is an external magnetic field along the z-axes the Hamiltonian is [44]

$$H = -J\sum_{ij} \vec{S}_i \cdot \vec{S}_j - \mu\vec{B}\cdot\sum_i \vec{S}_i. \tag{6}$$

Equation (5) describes an isotropic crystal. This can be further generalized for the anisotropic case. Suppose that x-axes and y-axes are equivalent but differ from the z-axes. Then, the extended Heisenberg Hamiltonian, given by Eq. (5), is [45]

$$H = -J\sum_{ij}\left(S_i^x S_j^x + S_i^y S_j^y\right) - K\sum_{ij} S_i^z S_j^z. \tag{7}$$

The Ising model, for which $J = 0$, and the ordinary Heisenberg model, for which $J = K$, are special cases of this Hamiltonian.

A further generalization is possible for two chain model or spin ladder [7], which is shown in Fig. 2. If both chains are in z-direction then all the spins are in xy-plane. It is obvious that there are two types of interactions. These are ferromagnetic-type interaction between the spins belonging to the same strand and antiferromagnetic-type interaction between the spins belonging to complementary strands. If the nearest neighbouring interactions are assumed then the most general Hamiltonian now reads [7]

$$H = \sum_n \Big[ -Jf_n\left(S_n^x S_{n+1}^x + S_n^y S_{n+1}^y\right) - Kf_n S_n^z S_{n+1}^z - J'f_n'\left(S_n'^x S_{n+1}'^x + S_n'^y S_{n+1}'^y\right)$$

$$-K'f_n' S_n'^z S_{n+1}'^z + \eta\left(S_n^x S_n'^x + S_n^y S_n'^y + S_n^z S_n'^z\right)\Big] \tag{8}$$

It is obvious that the last term multiplied by $\eta$ in Eq. (8) is the same as the trinome $T$ in Eq. (4). Note again that all the three terms in Eq. (4) have the same negative signs because of Eq. (3). Of course, $S_n^x$, $S_n'^x$, $S_n^y$ etc. in Eqs. (4) and (8) have different meanings. In Eq. (8) these are spin projections while in Eq. (4) these are the values of the base arrow projections defined by Eqs. (1) and (2). This analogy gives an idea to form the DNA Hamiltonian that is based on the generalized Heisenberg Hamiltonian [7]. Hence, if we use Eq. (8), include kinetic energy terms and assume that DNA is inhomogeneous we bring about the following DNA Hamiltonian

$$H = \sum_n \Big[ -Jf_n\left(S_n^x S_{n+1}^x + S_n^y S_{n+1}^y\right) - Kf_n S_n^z S_{n+1}^z - J'f_n'\left(S_n'^x S_{n+1}'^x + S_n'^y S_{n+1}'^y\right)$$

$$-K'f_n' S_n'^z S_{n+1}'^z + \eta\left(-S_n^x S_n'^x - S_n^y S_n'^y\right) + \mu S_n^z S_n'^z + \frac{I}{2}\dot{\Phi}_n^2 + \frac{I'}{2}\dot{\Phi}_n'^2 + \frac{I_\theta}{2}\dot{\theta}_n^2 + \frac{I_\theta'}{2}\dot{\theta}_n'^2 \Big] \tag{9}$$

where $I$, $I'$ etc. are the appropriate momenta of inertia of a single nucleotide. Note that the products $S_n^x S_n'^x$ and $S_n^y S_n'^y$ are negative, as was explained above. The inhomogeneity is involved in functions $f_n$ and $f_n'$, depending on $n$.

The terms including the parameters $J$, $K$, $J'$ and $K'$ describe the strong covalent bonds between the neighbouring nucleotides while those containing $\eta$ and $\mu$ describe the weak hydrogen bonds within the base pairs. We will see later that the term with $\mu$ will disappear and we should discuss the sign of the term containing $\eta$ only. Keeping in mind Eq. (3) we conclude that its overall sign is determined by the sign of the parameter $\eta$. If $S_n^x$, $S_n'^x$ etc. were real spins then the positive sign would indicate their antiparallel positions and vice versa ($\eta < 0$, spins parallel). Hence, according to Fig. 2, we may conclude that only positive $\eta$ has physical meaning. To be more precise, we consider the Heisenberg model of an isotropic spin ladder, Fig. 2, with two ferromagnetic legs coupled antiferromagnetically [46]. This means that $J$ and $J'$ are positive as corresponding energies are negative. On the other hand, the antiferromagnetic coupling energy, i.e. the term with the parameter $\eta$, should be positive. This is why we may expect that only positive $\eta$ has physical meaning. However, $S_n^x$, $S_n'^x$ etc. describe DNA geometry, as was shown by Eqs. (1) and (2). These are not real spins and, consequently, we do not want to exclude any value of $\eta$ for now. This point will be discussed later.

The Hamiltonian (9) can be rewritten in the limits of the plane-base rotator ($\theta_n = \theta_n' = \pi/2$). Hence, according to Eqs. (1), (2) and (9) we easily obtain

$$H = \sum_n \left[ \frac{I}{2}\dot{\Phi}_n^2 + \frac{I'}{2}\dot{\Phi}_n'^2 - Jf_n \cos(\Phi_{n+1} - \Phi_n) - J'f_n' \cos(\Phi_{n+1}' - \Phi_n') + \eta \cos(\Phi_n - \Phi_n') \right]. \quad (10)$$

A next step is to write down Hamilton's equations of motion for the generalized coordinates

$$q = \Phi_n, \qquad p = I\dot{\Phi}_n, \qquad q' = \Phi_n', \qquad p' = I'\dot{\Phi}_n'. \quad (11)$$

Also, we use a continuum approximation [7]

$$\Phi_n(t) \to \Phi(z,t), \quad \Phi_n'(t) \to \Phi'(z,t), \quad f_n \to f(z), \quad f_n' \to f'(z) \quad (12)$$

which will be justified for the specific artifficial, in vitro conditions, proposed in the next section. This brings about the following expansion

$$\Phi_{n\pm 1} = \Phi(z,t) \pm l\frac{\partial \Phi}{\partial z} + \frac{l^2}{2!}\frac{\partial^2 \Phi}{\partial z^2} \pm \ldots, \quad (13)$$

where the lattice parameter $l = 3.4 \,\text{Å}$ is the distance between the neighbouring nucleotides along the strands. Of course, we can write down the expansions for $\Phi_{n\pm 1}'$, $\Phi_{n\pm 1}'$, $f_{n\pm 1}$ and $f_{n\pm 1}'$ according to Eq. (13). From Eqs. (10)-(13) we can straightforwardly obtain, up to $O(l^2)$, the following equations of motion

$$I\Phi_{tt} = Jl^2[f(z)\Phi_{zz} + f_z\Phi_z] + \eta \sin(\Phi - \Phi') \quad (14a)$$
$$I'\Phi_{tt}' = J'l^2[f'(z)\Phi_{zz}' + f_z'\Phi_z'] - \eta \sin(\Phi - \Phi'). \quad (14b)$$

The suffices $t$ and $z$ represent respective partial derivatives. The two complementary bases rotate in opposite directions, as indicated in Fig. 1. Hence, we can assume that $\Phi' = -\Phi$. Also, we safely assume [7] $I = I'$, $J = J'$ and $f = f'$ and, by subtracting Eqs. (14), we easily obtain

$$I\psi_{tt} = Jl^2 f(z)\psi_{zz} + Jl^2 f_z(z)\psi_z + 2\eta \sin\psi, \qquad (15)$$

where $\psi \equiv \psi(z,t) = 2\Phi$. For the considered homogeneous strands the function $f(z) = const$ and Eq. (15) becomes

$$\psi_{tt} - c_0^2 \psi_{zz} + \omega_0^2 \sin\psi = 0, \qquad (16)$$

where

$$c_0^2 = \frac{Jl^2 f}{I}, \qquad \omega_0^2 = -\frac{2\eta}{I}. \qquad (17)$$

Notice that there are three unknown parameters in Eq. (16) and these are $Jf$, $I$ and $\eta$. Equation (16) is the well known, integrable, sine-Gordon equation. Its particular, but very important solution, is [47]

$$\psi = 4\tan^{-1}\exp[\pm \kappa(z - vt)], \qquad (18)$$

where

$$\kappa = \frac{\omega_0}{c_0}\left(1 - \frac{v^2}{c_0^2}\right)^{-\frac{1}{2}}, \qquad v < c_0. \qquad (19)$$

These are well known kink (+) and antikink (-) solitons. Of course, all this makes sense for $\eta < 0$.

Obviously, for the positive parameter $\eta$ we obtain the equation

$$\psi_{tt} - c_0^2 \psi_{zz} - \omega_0^2 \sin\psi = 0, \qquad (20)$$

where

$$c_0^2 = \frac{Jl^2 f}{I}, \qquad \omega_0^2 = \frac{2\eta}{I}. \qquad (21)$$

Following the same procedure [47] yielding the solution (18) we can easily see that Eq. (18) holds again, i.e. represents the solution of Eq. (20), with new wave vector

$$\kappa = \frac{\omega_0}{c_0}\left(\frac{v^2}{c_0^2} - 1\right)^{-\frac{1}{2}}, \qquad v > c_0. \tag{22}$$

In what follows, we pay attention to the velocities $v$ and $c_0$. We offer the experimental proposals explaining how these speeds can be determined. A crucial point is to know which one is bigger, $v$ or $c_0$, as this could indicate the type of interaction. In other words, this knowledge can determine the sign of the parameter $\eta$.

Also, we explain how to experimentally determine the solitonic width $\Lambda$. Hence, if $v$, $c_0$ and $\kappa$ are known then we can compare the two basic interactions in DNA molecule. Namely, according to Eqs. (19), (21) and (22) we see that the wave number reads

$$\kappa \equiv \frac{2\pi}{\Lambda} \propto \sqrt{\frac{2|\eta|}{Jf}}. \tag{23}$$

Therefore, the wave number $\kappa$ depends on the ratio of the hydrogen bond interaction and the covalent bond interactions $\eta$ and $Jf$, as was explained above. Even more, since we have

$$\kappa = \omega_0 F(v, c_0), \qquad \omega_0 \propto \sqrt{\frac{|\eta|}{I}}, \tag{24}$$

where $F(v, c_0)$ is an obvious function of $v$ and $c_0$, determined by either Eq. (19) or Eq. (22), we can determine the ratio $\eta/I$. Hence, if $I$ were experimentally determined, or, at least, estimated, which is not a topic of this paper, the value of the crucial parameter $\eta$ would be known or estimated. Of course, this would bring about the known or estimated value of the parameter $Jf$.

Finally, for $\eta = 0$, Eqs. (16) and (20) reduce to an ordinary linear wave equation which solution is a plane wave with a speed $c_0$. Hence, $v$ is the solitonic velocity in DS-DNA while $c_0$ is the plane wave velocity in SS-DNA. In order to exclude the role of base pairs interactions which is responsible for appearance of nonlinear sine-term in Eqs. (16) and (20) one would try with aligned SS-DNA chain oriented normally with the homogeneous electric field aimed to arrange base arrows parallely with field's vector.

It might be interesting to point out that some values of $I$ can be found in literature, but they differ remarkably. The reported values are $5 \cdot 10^{-44}\,\text{kgm}^2$ [48] and $34 \cdot 10^{-44}\,\text{kgm}^2$ [49] which means that further research should be carried out in order to find a reliable value.

**Solitonic speed**

DNA is one of a few polymers that can be twisted, wound or unwound. Nature has evolved a vast kit of enzymes to paste, unwound or repair DNA molecules. Some are used to modify DNA ends for anchorage to appropriate surface enabling it to manipulate, stretch and twist single DNA.

The well known "twisters" are RNA-polymerase creating large torques in DNA and DNA gyrase with pretty lower peak torques. The energy spent for these twists originates from ATP hydrolysis.

In the study of DNA molecules it was found that the magnetic-tweezer technique is very efficient. Briefly, it enables the stretching and twisting a single DNA while it is anchored at one end to a substrate surface and, at the other end, to a magnetic bead whose diameter is of the order of a $\mu m$.

A small magnet, whose position and rotation can be controlled, is used to pull and rotate above micro bid, thus stretching and unbinding DNA. We propose a principle for the experimental strategy for the homogeneous totally unwound chain of DS-DNA being excited to make base rotations which could be identified and detected after passing the known distance.

The unwinding of DNA is formed by the rotation of the magnetic micro bid (MMB), Fig. 3, on one (left) side of stretched B-DNA. Close to the left end of the chain a small nanomagnet (NM) is intercalated within base pairs. The local twisting torque is formed by the rotation of intercalated nanomagnet bead (NM) in terms of applied oscillating magnetic field. Close to the anchored right end, the fluorescent detector should be used. This is the intercalated appropriate fluorescent labelling molecule. The compound ethidium bromide (ETB) when free to change its conformation in solution, has very little fluorescence. Its fluorescence is greatly enhanced when it binds to DNA. The hydrophobic environment found between the base pairs is believed to be responsible for ETB's intense fluorescence after binding with DNA. By moving into this hydrophobic area and being away from the solvent, the ETB cation is forced to shed any water molecules that were associated with it. As water is highly efficient fluorescent quencher, the removal of these water molecules allows the ETB to fluoresce. When the labelled DNA segment is subjected under torsional rotations of incoming kink wave, the accompanying movement of intercalated ETB can be detected by measuring its dipole orientation by analysing polarized fluorescence.

The second option for kink detection is a removed protein detector. This is the removal of DNA-bound proteins by DNA twisting provided by the incoming kink. Sarkar and Marko [50] had published the model in which the conditions for threshold twisting torques are considered. For the case of nucleosome octamer bound to DNA they found the critical torque to be about $9k_BT$ (nine thermal energy). If the passing kink has enough energy it could remove above protein thus getting detected, since the energy of the kink can be tuned by the amplitude of twisting magnet torque (on the left of Fig. 3) responsible for its excitation.

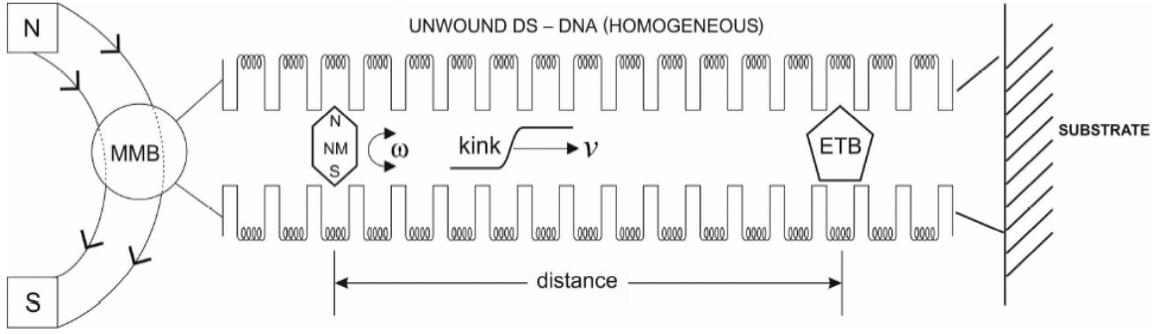

**Fig. 3** Experimental set up for measurement of kink velocity $v$

Therefore, it is explained how the kink can be excited and how its arrival can be detected. The distance between these two locations is measurable. If the instants when these two successive events occur are determined the solitonic speed $v$ can easily be calculated.

Next important step is the determination of speed $c_0$. It was pointed out that Eqs. (16) and (20) become the ordinary linear wave equation for $\omega_0 = 0$. Its solution is the progressing plain-wave and its velocity is just $c_0$. It is apparent that the condition $\omega_0 = 0$ follows from the restriction $\eta = 0$. It was also pointed out that the parameter $\eta$ describes the interaction between the two strands, as can be seen from Eq. (8). Therefore, $\omega_0 = 0$ means that the two strands do not interact and DNA molecule is separated in two independent single strands. This implies that $c_0$ is the velocity of the plain acoustic wave characterizing the single strand. We now propose a possible assay in order to asses the velocity $c_0$ along the SS chain of DNA. To achieve this goal the SS chain is stretched to be strait by two laser tweezers, applied at opposite ends, and denoted by LT (light trap), as shown in Fig. 4. Two polystyrene beads (PSB) are used to attach DNA to LTs. The DNA bases are oriented paralelly with an applied homogeneous electric field $E$. This field plays an analogous role as gravitational field does play for the collection of coupled physical pendula [51]. The twisting torque is formed by the rotation of a similar intercalated NM close to the left end, as before, in terms of applied oscillating magnetic field of small amplitude.

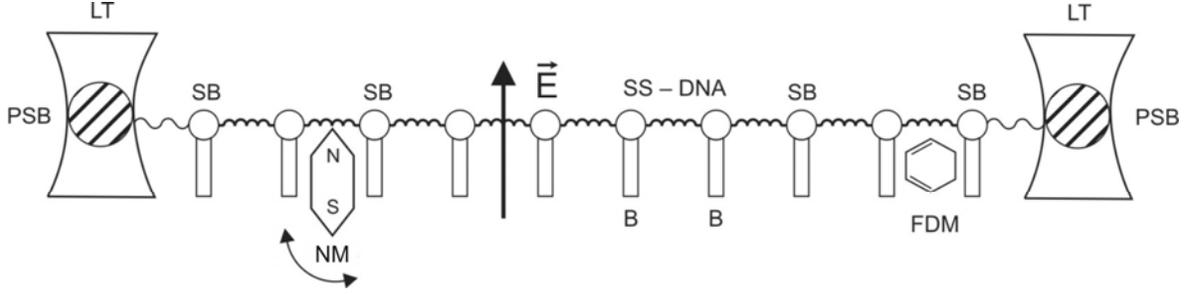

**Fig. 4** The excitation and detection of the kink-soliton in SS-DNA (Abbreviations SP and B denote sugar-phosphate and base.)

If small amplitude wave is being excited by NM oscillation its passage could be identified by the rotation of the direction of emitted fluorescent photons from fluorescent dye molecule (FDM) intercalated close to the right end of SS-DNA. It could be monitored by fluorescent microscopy capable to detect its oscillations caused by incoming wave from the left. This gives the opportunity to measure the time delay between wave initiation on the left, and its detection on the right hand side.

**Solitonic width**

Let us study the parameters of a soliton

$$F(x,t) = \frac{12}{\pi} \tan^{-1}\left[\exp\left(-\frac{x-vt}{L}\right)\right]. \tag{25}$$

A common factor 4 existing in Eq. (18) is multiplied by $3/\pi$ to obtain an amplitude which is exactly six. The function (25) is shown in Fig. 5, for $x = 120$, $v = 2$ and $L = 4$.

The experimental wave detector should detect the beginning and the end of the soliton. These are the points A and B in Fig. 5 with the coordinates $F_A$ and $F_B$ close to zero and six respectively.

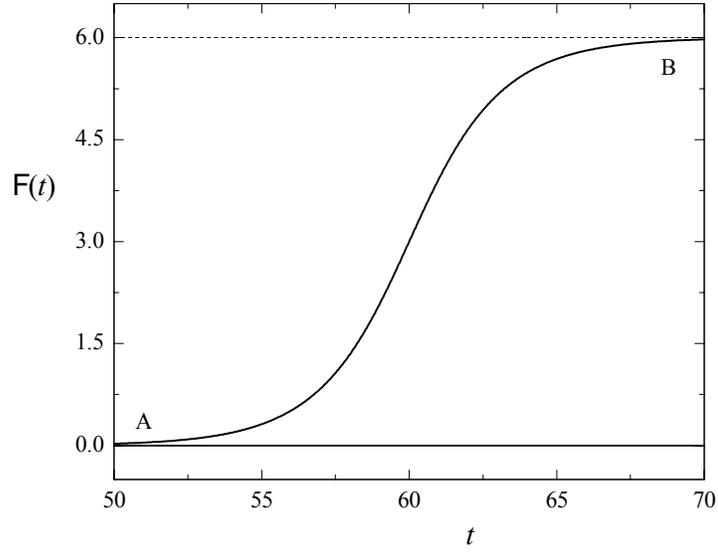

**Fig. 5** A soliton $F(x,t) = \dfrac{12}{\pi} \tan^{-1}\left[\exp\left(-\dfrac{x-vt}{L}\right)\right]$ for $x = 120$, $v = 2$, $L = 4$

It is important to notice that $L$ in Eq. (25) is not the solitonic width. If $L$ were the width then the distance $L$ would be passed during time $\Delta t = L/v$, i.e. between the moments $t_1 = \dfrac{x}{v} - \dfrac{L}{2v}$ and $t_2 = \dfrac{x}{v} + \dfrac{L}{2v}$ where $\dfrac{x}{v} = 60$ is the centre of the soliton as can be seen from Fig. 5. The appropriate values of $F$ are $F(t_1) = 2.1$ and $F(t_2) = 3.9$, and these are not the values $F_A$ and $F_B$ that should be experimentaly detected. Therefore, it is obvious that $L$, corresponding to the time instant $\Delta t \equiv t_2 - t_1$, is much smaller than the solitonic width.

There is a better definition for the solitonic width $\Lambda$. Equation (25) suggests that equality $2\pi/\Lambda = 1/L$ holds, and, therefore,

$$\Lambda = 2\pi L. \tag{26}$$

Hence, we are looking for the points at the moments:

$$t_A = \dfrac{x}{v} - \dfrac{\pi L}{v}, \qquad t_B = \dfrac{x}{v} + \dfrac{\pi L}{v}. \tag{27}$$

The appropriate values of $F$, for this particular example are, $F_A = 0.16$ and $F_B = 5.84$. Hence, these are values representing the beginning and the end of the soliton that can be experimentaly determined.

Therefore, the detector can determine the instants $t_A$ and $t_B$, i.e. the interval $\Delta t = t_B - t_A$. If the speed $v$ is known we can calculate the solitonic width as $\Lambda = v\,\Delta t$ and, finally, the value of $L$. This, practically, means that we can determine the ratio $\eta/Jf$, as explained above. Of course, this is very important as this represents the ratio between the hydrogen-bond interaction and the covalent interaction in DNA.

**Conclusion**

Biophysics has developed a couple of models describing nonlinear dynamics of DNA molecules. According to the PBR model it is expected that sine-Gordon kink and antikink solitons could appear in DS-DNA chain.

The experiments that are suggested in this paper are aimed to test this theoretical prediction. They are primarily intended to measure the solitonic speed and its effective width. To test this basic theory only the homogeneous DS-DNA segment should be prepared and used for the experiment. If for such artificial specimen in vitro conditions experiment exhibits the existence of kink solitons it could be used to compare their speed $v$ and the speed $c_0$ of the linear waves propagating along SS-DNA chain of the same structure. One should keep in mind that we are primarily looking for the ratio of the two velocities $v/c_0$ as this determines the sign of the parameter $\eta$, which has the basic consequence for the BPR model.

It is apparent that the proposed theoretical model is not completely realistic as damping forces are not taken into consideration, but we expect that for a DNA chain of moderate length the damping should not kill the excited kink before it is being detected. The next step in improving this simple PBR model should be the incorporation of the viscous term in the equation of motion which is the topic of forthcoming paper.

**Acknowledgements** We want to thank to Dejan Milutinović for his technical help. S. Zdravković and M. Daniel gratefully acknowledge the hospitality of the Abdus Salam International Centre for Theoretical Physics, Trieste, Italy, where most of this work was done. S. Zdravković wants to thank to Kavitha Louis for very useful discussions. This research was supported by funds from Serbian Ministry of Sciences, grants 142034G, III45010 and OI171009.